\documentclass[12pt]{article}
\usepackage{amsfonts,amsmath}
\usepackage{amssymb}
\usepackage{dsfont}
\usepackage{mathtools}
\usepackage{xfrac}

\newcommand {\RN} [1] {\uppercase \expandafter {\romannumeral #1}}

\makeatletter \@addtoreset{equation}{section} \makeatother
\newcommand{\dr}{{{\rm d}}}
\newcommand{\DD}{{{\mathbb{D}}}}
\newcommand{\rhs}{$r.h.s.$}

\begin{document}
\begin{flushright}
FIAN/TD/09-25\\
\end{flushright}

\vspace{0.5cm}
\begin{center}
{\large\bf Differential Contracting Homotopy in the Linearized 3d Higher-Spin Theory}

\vspace{1 cm}

\textbf{M.A.~Vasiliev${}^{1,2}$ and V.A.~Vereitin${}^{1,2}$}\\

\vspace{1 cm}

\vspace{0.5 cm}
		\textbf{}\textbf{}\\
		\vspace{0.5cm}
		\textit{${}^1$ I.E. Tamm Department of Theoretical Physics,
			Lebedev Physical Institute,}\\
		\textit{ Leninsky prospect 53, 119991, Moscow, Russia}\\
		
		\vspace{0.7 cm} \textit{
			${}^2$ Moscow Institute of Physics and Technology,\\
			Institutsky lane 9, 141700, Dolgoprudny, Moscow region, Russia
		}

\par\end{center}

\begin{center}
\vspace{0.6cm}

\par\end{center}

\vspace{0.4cm}

\begin{abstract}
In this paper, the recently developed differential homotopy approach is applied to the problem of disentangling dynamical and topological fields of the $3d$ higher-spin gauge theory at the linear level. This formalism allows us to reproduce all known disentangling solutions in a unified form, including both the solutions obtained previously within the shifted homotopy approach in \cite{Korybut:2022kdx} and those derived by hand in \cite{Vasiliev:1992ix}, as well as other solutions including those associated with the cohomology of the background covariant derivative $D_0$.
Also, within the differential homotopy framework, an alternative way of derivation of disentangled equations through a non-conventional solution for the field $S_1$ is suggested.
These results are crucial for further investigation of nonlinear corrections to HS equations in $AdS_3$.
\end{abstract}

\newpage
\tableofcontents

\newpage

\section{Introduction}\label{Xsec1-1} \label{Intro}

Nonlinear higher-spin (HS) equations in $AdS_3$ can be formulated using the system of \cite{Prokushkin:1998bq} formulated in terms of generating functions for the HS one-form gauge fields $\omega$ and zero-form matter fields $C$. Following \cite{Vasiliev:1992ix,Prokushkin:1998bq,Vasiliev:1995dn}, the nonlinear field equations can be put into the unfolded form of a perturbative expansion in powers of $C$,
\begin{align}
&\dr_x \omega + \omega \ast \omega = \Upsilon(\omega,\omega,C) + \Upsilon(\omega,\omega,C,C) + \ldots,\label{1form}\\
&\dr_x C + \omega \ast C - C \ast \omega = \Upsilon(\omega,C,C) + \Upsilon(\omega,C,C,C) + \ldots,\label{0form}
\end{align}
where $\dr_x := dx^\nu \dfrac{\partial}{\partial x^\nu}$ is the space-time de Rham derivative.
Throughout this paper, wedge product symbol for differential forms is implicit.

As is well known, $3d$ massless HS gauge fields $\omega$ do not propagate \cite{Blencowe:1988gj}. However, the theory involves propagating \textit{dynamical} matter fields in the sector of zero-forms $C$. The latter also contain \textit{topological} fields, which do not propagate, belonging to finite-dimensional modules of the isometry algebra. Topological fields appear on the \rhs's of \eqref{1form}, \eqref{0form}, and the two types of fields may source each other. This phenomenon is called \textit{entanglement}\footnote{Here, the term entanglement implies an unwanted mixing of different types of fields, having no relation with quantum entanglement.}.
For the analysis of the theory, it is desirable to disentangle the equations for topological and dynamical fields. This problem was originally solved at the linearized level in \cite{Vasiliev:1992ix} by an appropriate field redefinition. Later on, using the \textit{shifted homotopy} method \cite{Didenko:2018fgx}, a two-parameter family of disentangling solutions to the linearized equations was found \cite{Korybut:2022kdx}. However, this family does not describe all possible disentangling solutions. In particular, it does not reproduce the solution of \cite{Vasiliev:1992ix}.

In \cite{Vasiliev:2023yzx}, a new \textit{differential homotopy} method was proposed with the parameters of the shifted homotopy approach treated as integration variables over a polyhedron in a multidimensional manifold $\mathcal{M}$.
This approach is more general, yielding solutions in the form of certain \textit{fundamental Ansatz}, which in particular trivializes Schouten identities, a major technical challenge in the shifted homotopy framework.
The aim of this paper is to apply the differential homotopy approach to the entanglement problem at the linear level of $3d$ HS theory. This formalism allows us to reproduce all known disentangling solutions in a unified form of the fundamental Ansatz and derive the general disentangling solution unifying those of \cite{Vasiliev:1992ix} and \cite{Korybut:2022kdx}.

The rest of the paper is organized as follows. The remainder of Section {\ref{Intro}} provides a general introduction to $3d$ HS theory, including the current state of the entanglement problem. Section {\ref{Diffhom}} sketches the differential homotopy technique. Section {\ref{Linearlevel}} presents a general scheme of solving linearized equations via differential homotopies. Section {\ref{DisentanglingdW1}} shows the construction of a complete family of disentangling corrections to the equation for one-forms. Section {\ref{DisentanglingS1}} demonstrates an alternative construction of disentangling solutions by choosing a non-standard one-form $S_1$.
Conclusions and future perspectives are discussed in Section {\ref{Conclusion}}.

\subsection{Higher spins in $AdS_3$}\label{Xsec2-1.1}

In Cartan gravity, the geometry of $AdS_n$ can be described in terms of an
$o(n-1,2)$ flat connection $W_{gr}$,
\begin{equation}
 \dr_x W_{gr} + W_{gr} W_{gr} = 0.
\label{Xeqn3-3}
\end{equation}
The $AdS_3$ connection with the symmetry algebra $o(2,2) \sim sp(2) \oplus sp(2)$ and local coordinates $x^{\nu}$ ($\nu = 0,1,2$) decomposes into the spin-connection $\omega_0^{\alpha \beta}$ and the dreibein $h_0^{\alpha \beta}$,
\begin{equation}
\omega_0^{\alpha \beta}(x) = \omega_{0\nu}^{\alpha \beta}(x) dx^{\nu}, \quad
h_0^{\alpha \beta}(x) = h_{0 \nu}^{\alpha \beta}(x) dx^{\nu} ,
\label{Xeqn4-4}
\end{equation}
with $sp(2)$ spinor indices $\alpha, \beta = 1,2$ raised and lowered by the symplectic form $\epsilon^{\alpha \beta}$,
\begin{equation}\label{indexrule}
\begin{array}{ll}
A_{\beta} = A^{\alpha} \epsilon_{\alpha \beta},
\quad
A^{\alpha} = \epsilon^{\alpha \beta} A_{\beta},
\quad
\epsilon_{\alpha \beta} = -\epsilon_{\beta \alpha},
\quad
\epsilon_{\alpha \beta} \epsilon^{\alpha \gamma} = \delta_{\beta}^{\gamma}.
\end{array}
\end{equation}

The $AdS_3$ geometry is characterized by zero-curvature and zero-torsion conditions,
\begin{equation}\label{curvspinor}
\dr_x \omega_0^{\alpha \beta} + \omega_0^{\alpha}{}_{\gamma} \omega_0^{\beta \gamma} + \lambda^{2} h_0^{\alpha}{}_{\gamma} h_0^{\beta \gamma} = 0,
\end{equation}
\begin{equation}\label{torsspinor}
\dr_x h_0^{\alpha \beta} + \omega_0^{\alpha}{}_{\gamma} h_0^{\beta \gamma} + h_0^{\alpha}{}_{\gamma} \omega_0^{\beta \gamma} = 0,
\end{equation}
where $\lambda^2$ represents the cosmological constant.

It is convenient to introduce spinor variables $y_\alpha$ that commute under the usual juxtaposition product but satisfy the Heisenberg commutation relation $[y_{\alpha}, y_{\beta}]_* = 2i\epsilon_{\alpha \beta}$ with respect to the star product
\begin{equation}
f(y) * g(y) = \int d^{2} u \, d^{2} v \, f(y+u) g(y+v) e^{i u_{\alpha} v^{\alpha}}.
\label{Xeqn8-8}
\end{equation}

The background $AdS_3$ connection can be represented as
\begin{equation}\label{AdSback}
W_0(y) = \frac{1}{4i}\omega_0^{\alpha\beta} y_\alpha y_\beta + \frac{\lambda}{4i} \psi_1 h_0^{\alpha\beta} y_\alpha y_\beta = \omega_0(y) + \lambda \psi_1 h_0(y),
\end{equation}
satisfying the equation
$
\dr_x W_0 + W_0 * W_0 = 0,
$
which is equivalent to \eqref{curvspinor}-\eqref{torsspinor}. Following \cite{Prokushkin:1998bq,Vasiliev:1995dn}, we introduce the Clifford elements $\psi_i$, $\{\psi_{i}, \psi_{j}\} = 2 \delta_{i j}$, which commute with the spinor variables $y_\alpha$ and the differential forms. Their role is clarified below.

The unfolded form of the free equations \cite{Prokushkin:1998bq,Vasiliev:1992gr} can be written in terms of the
generating function, constructed as a formal series in spinor variables
\begin{equation}\label{Cgen}
C(y; \psi; k|x) = \sum_{A, B, C=0}^{1} \sum_{n=0}^{\infty} \frac{1}{n!} \lambda^{-\left[\frac{n}{2}\right]} C_{\alpha_{1} \ldots \alpha_{n}}^{A B C}(x) k^{A} \psi_{1}^{B} \psi_{2}^{C} y^{\alpha_{1}} \ldots y^{\alpha_{n}}.
\end{equation}
Here, $k$ is the Klein operator, anticommuting with spinor variables, $k y_\alpha = - y_\alpha k$, and commuting with all other symbols.

Introducing the covariant derivative in $AdS_3$ as
\begin{equation}\label{D0}
D_{0} P = \dr_x P + W_{0} * P - (-1){}^{p} P * W_{0}
\end{equation}
for a degree-$p$ differential form $P$, the matter field equation takes the form of a covariant constancy condition,
\begin{equation} \label{D0C}
D_{0} C = 0.
\end{equation}

In the expansion of this field in powers of $\psi_2$,
\begin{equation}\label{C^dt}
C\left(y ; \psi ; k\right) = C^{\text{top}}\left(y ; \psi_{1} ; k\right) + C^{\text{dyn}}\left(y ; \psi_{1} ; k\right) \psi_2 ,
\end{equation}
the first and second terms are called topological and dynamical, respectively.
The origin of this terminology is as follows. Introducing a Lorentz-covariant derivative as
$
D_L C := \dr_x C + [\omega_0, C]_* ,
$
Eq. {\eqref{D0C}} splits into two independent subsystems:
\begin{equation}
D_L C^{\text{top}} = -\lambda \psi_1 \left[ h_0 , C^{\text{top}} \right]_* , \quad
D_L C^{\text{dyn}} = -\lambda \psi_1 \left\{ h_0 , C^{\text{dyn}} \right\}_* .
\label{Xeqn14-14}
\end{equation}
Using the decomposition~\eqref{Cgen}, one obtains
\begin{equation}\label{DLCgen}
\begin{split}
D_L C_{\alpha(n)}^{\text{top}} = 4 i \lambda n h_{0\,\alpha}{}^{\beta} C_{\beta \alpha(n-1)}^{\text{top}} , \quad
D_L C_{\alpha(n)}^{\text{dyn}} = \frac{\psi_1}{2i} \left[ h_0^{\gamma \delta} C^{\text{dyn}}_{\gamma \delta \alpha(n)}
- \lambda^2 n(n-1)\, h_{0\,\alpha \alpha} C^{\text{dyn}}_{\alpha(n-2)} \right] .
\end{split}
\end{equation}
Here, the left equation relates a finite number of component fields at each $n$, while the right one generates two infinite chains of equations with an unbounded (odd or even) numbers of indices. The \rhs of $D_L C_{\alpha(n)}^{\text{dyn}}$ corresponds to the free field equations in $AdS_3$, which reduce to the Klein-Gordon and Dirac equations at $n=2$ and $n=1$, respectively,
\begin{equation}
D_L^{\mu} D_{L \mu} C^{\text{dyn}} = \frac{3}{2} \lambda^2C^{\text{dyn}} , \quad
{h_0}^\nu{}_\alpha{}^\beta D_{L \nu} C^{\text{dyn}}_\beta = 0 .
\label{Xeqn16-16}
\end{equation}

In $AdS_3$ HS theory, there are only two dynamical fields with spins $0$ and $\frac{1}{2}$. The other components of $C_{\alpha(n)}^{\text{dyn}}$ are expressed by \eqref{DLCgen} via their higher space-time derivatives. Topological fields, as emphasized in \cite{Didenko:2015pjo}, can be interpreted as coupling constants in the HS theory. The elements $\psi_i$ allow one to write both dynamical and topological equations in a unified form, as in \eqref{D0C}. However, this leads to the entanglement problem discussed in Section {\ref{Entprob}}.

\subsection{Higher-spin nonlinear equations}\label{Xsec3-1.2}

The nonlinear generating system for HS equations of \cite{Prokushkin:1998bq} can be equivalently formulated as
\begin{equation}\label{PV1}
\dr_x W + W * W = 0 ,
\end{equation}
\begin{equation}\label{PV2}
\dr_x B + [W, B]_* = 0,
\end{equation}
\begin{equation}\label{PV3}
\dr_x S + \{W, S\}_* = 0,
\end{equation}
\begin{equation}\label{PV4}
S * S = i \left( \theta^{\alpha} \theta_{\alpha} + B * \gamma \right) ,
\end{equation}
\begin{equation}\label{PV5}
[S, B]_* = 0 .
\end{equation}
Apart from space-time coordinates $x^\nu$ with the de Rham derivative $\dr_{x} := d x^\nu \dfrac{\partial}{\partial x^\nu}$, the master fields $W(z; y; \psi; k | x)$, $B(z; y; \psi; k | x)$, and $S(z; \theta; y; \psi; k | x)$ depend on the spinor variables $z_\alpha, y_\alpha$ ($\alpha = 1, 2$), the Clifford elements $\psi_i$, and the Klein operator $k$.

There are two types of anticommuting differentials in \eqref{PV1}-\eqref{PV5}, namely space-time $d x^\nu$ and spinor $\theta^\alpha := dz^\alpha$. The master fields are graded differently with respect to these differentials: $W = W_{n} d x^\nu$ and $S = S_{\alpha} \theta^{\alpha}$ are one-forms in $d x^\nu$ and $\theta^\alpha$, respectively, while $B$ is a zero-form. The two-form operator $\gamma$ in \eqref{PV4} is defined as $\gamma = \theta^{\alpha} \theta_{\alpha} \exp \left(i z_{\alpha} y^{\alpha}\right) k$.

The (anti)commutation relations are
\begin{equation}\label{comm}
\begin{gathered}
{[y_{\alpha}, y_{\beta}]=[z_{\alpha}, z_{\beta}]=[z_{\alpha}, y_{\beta}]=0}, \\
\{d x_\mu, d x_\nu\}=\{\theta_{\alpha}, \theta_{\beta}\}=\{\theta_{\alpha}, d x_\nu\}=0 ,\\
\{k, y_{\alpha}\}=\{k, z_{\alpha}\}=\{k, \theta_{\alpha}\}=0, \quad [k, dx_\nu] = 0, \quad k^{2}=1, \\
{[\psi_{i}, y_{\alpha}]=[\psi_{i}, z_{\alpha}]=[\psi_{i}, dx_\nu]=[\psi_{i}, \theta_\alpha]=[\psi_{i}, k]=0, \quad\{\psi_{i}, \psi_{j}\}=2 \delta_{i j}}.
\end{gathered}
\end{equation}

The extended star product of formal power series in spinor variables $y_\alpha$ and $z_\alpha$ is defined as
\begin{equation}\label{*-prod}
f(z, y) * g(z, y) := \int d^{2} u \, d^{2} v f(z+u, y+u) g(z-v, y+v) e^{i u_{\alpha} v^{\alpha}} .
\end{equation}
This star product generates the commutation relations $[y_\alpha, y_\beta]_*= -[z_\alpha, z_\beta]_* = 2i \epsilon_{\alpha \beta}$, $[y_\alpha, z_\beta]_*=0$.

Due to the Schouten identity $\theta^3 = 0$, the operator $\gamma$ in \eqref{PV4} turns out to be central,
\begin{equation}\label{gamma}
\gamma = \theta^{\alpha} \theta_{\alpha} \exp \left(i z_{\alpha} y^{\alpha}\right) k ,
\quad
\gamma * f = f * \gamma, \quad \forall f = f(z; y; k; \psi; \theta) .
\end{equation}
To see this, one has to take into account that $\exp \left(i z_{\alpha} y^{\alpha}\right)$ plays the role of the inner Klein operator,
\begin{equation}
 e^{i z_\alpha y^\alpha} * f(z, y) = f(-z, -y) * e^{i z_\alpha y^\alpha} .
\label{Xeqn25-25}
\end{equation}

Following \cite{Vasiliev:1992ix}, to analyze Eqs. {\eqref{PV1}}-\eqref{PV5} perturbatively, one starts with the vacuum solution
\begin{equation}\label{vacsol}
 B_0 = 0, \quad
 S_0 = \theta^\alpha z_\alpha, \quad
 W_0 = \frac{1}{4i} (\omega_0 + \lambda \psi_1 h_0){}^{\alpha\beta} y_\alpha y_\beta,
\end{equation}
where $W_0$ is the background $AdS_3$ connection \eqref{AdSback}. The star-commutator of any formal function $f(z,y)$ with $S_0$ acts as the $z_\alpha$ exterior derivative
\begin{equation}\label{S0dz}
 [S_0, f]_* = -2i \dr_z f, \qquad \dr_z := \theta^\alpha \dfrac{\partial }{\partial z^\alpha} .
\end{equation}

In the next order, from $\left[S_{0}, B_{1}\right]_{*} = 0$ it follows that $B_{1} = C(y ; \psi; k \mid x)$, with $C(y ; \psi; k \mid x)$ obeying \eqref{D0C} by virtue of \eqref{PV2}.

\subsection{Linear entanglement problem}\label{Xsec4-1.3} \label{Entprob}

At the first perturbative level we set $B = C$, $W = W_{0} + W_{1}$, and $S = S_{0} + S_{1}$. From the vacuum solution \eqref{vacsol} and the property \eqref{S0dz}, Eq. {\eqref{PV5}} implies that $C$ is $z$-independent, while \eqref{PV2} gives the field Eq. {\eqref{D0C}}. Eqs. {\eqref{PV3}} and \eqref{PV4} yield
\begin{equation}\label{PV34lin}
 \{S_0, S_1\}_* = i C * \gamma, \qquad D_0 S_1 + \{S_0, W_1\}_* = 0 \, .
\end{equation}

As a consequence of \eqref{S0dz}, Eqs. {\eqref{PV34lin}} have the form 
\begin{equation} \label{dzf}
\dr_{z} f(z ; y ; \theta; \psi; k) = J(z ; y ; \theta; \psi; k), \quad \text{where} \quad \dr_z J(z ; y ; \theta; \psi; k) = 0, \, J(z ; y ; 0; \psi; k) = 0.
\end{equation}
Due to the property of $S_0$ \eqref{S0dz}, the problems of the type \eqref{dzf} occur at every order of the perturbative analysis of Eqs. {\eqref{PV1}}-\eqref{PV5}. In \eqref{dzf}, these are formulated in a general way for formal power series in the spinor variables $y_\alpha$, $z_\alpha$, which may also depend on $\theta_\alpha, \psi_i$ and $k$.

These can be solved by virtue of the Poincaré formula
\begin{equation}
\label{dh}
\Delta_{0} J(z ; y ; \theta) := z^{\alpha} \frac{\partial}{\partial \theta^{\alpha}} \int_{0}^{1} \frac{d \tau}{\tau} \, J(\tau z ; y ; \tau \theta) , \qquad
f = \Delta_{0} J + \dr_{z} \epsilon + h ,
\end{equation}
where $h$ is in the $\dr_{z}$-cohomology.
Generally, the exact part $\dr_{z} \epsilon$ and the cohomology $h$ in \eqref{dh} describe solutions to the homogeneous Eq. {\eqref{dzf}} with $J = 0$. The freedom in $\epsilon$ is gauge, whereas $h(\omega, C)$ with HS fields $\omega(y ; k ; \psi \mid x)$ and $C(y ; k ; \psi \mid x)$ induces field redefinitions.

Though it may seem most natural, the conventional homotopy with $\epsilon = h = 0$ may lead to undesirable consequences. For example, the conventional homotopy in the second order of the perturbative expansion of the $4d$ theory leads to non-local results \cite{Giombi:2009wh,Boulanger:2015ova}. In the $3d$ HS theory under consideration, it entangles dynamical and topological fields even at the linear level \cite{Vasiliev:1992ix}.

Shifted homotopy operator results from the shift $z^\alpha \rightarrow z^\alpha + q^\alpha$ with some $z$-independent $q^\alpha$. Shifted contracting homotopy $\Delta_{q}$ and cohomology projector $h_{q}$ act as follows \cite{Gelfond:2018vmi}
\begin{equation}\label{Dq}
\Delta_{q} J(z ; y ; \theta):=\left(z^{\alpha}+q^{\alpha}\right) \frac{\partial}{\partial \theta^{\alpha}} \int_{0}^{1} \frac{d \tau}{\tau} J(\tau z-(1-\tau) q ; y ; \tau \theta), \quad
h_{q} f(z ; y;\theta)=f(-q ; y;0).
\end{equation}

The form of the resulting equations at $\epsilon = h = 0$ depends on the choice of $\Delta_{q}$. Transition from one contracting homotopy to another affects both $\epsilon$ and $h$. The problem is to identify a homotopy procedure that leads to disentangled field equations.

The star-exchange formulas from \cite{Didenko:2018fgx} allow one to move the homotopy operator through the $*$-symbol up to the $\gamma$ element \eqref{gamma}.
So, the conventional solution of \eqref{PV34lin} for $S_1$ can be represented in the form
\begin{equation}
S_1(z; \theta; y; \psi; k|x) = \Delta_0 \Big(-\frac{1}{2} C(y; \psi;k|x) * \gamma \Big) = -\frac{1}{2} C(y; \psi;k|x) * \Delta_p \gamma ,
\label{Xeqn32-32}
\end{equation}
where $p_\alpha$ denotes the derivative with respect to the total argument $\xi^\alpha$ of the field $C(\xi)$. The corresponding derivative of one-forms $\omega(\xi)$ is denoted $t_\alpha$,
\begin{equation}\label{dpt}
p_\alpha C(\xi) \equiv C(\xi) p_\alpha:=-i \frac{\partial}{\partial \xi^\alpha} C(\xi) , \quad t_\alpha \omega(\xi) \equiv \omega(\xi)t_\alpha := -i \frac{\partial}{\partial \xi^\alpha}\omega(\xi).
\end{equation}

This leads to the following $\dr_z$-differential equation on the linear part of $W(z;y;\psi_{1,2};k)$:
\begin{equation}\label{dZW1}
\dr_z W_1 = - \frac{1}{4i} \left[ W_0 * C * (\Delta_{p}-\Delta_{p+t})\gamma + C * W_0 * (\Delta_{p+t}-\Delta_{p+2t})\gamma \right] := J_W(y;z;\theta).
\end{equation}

The conventional homotopy yields a solution of the form:
\begin{equation}\label{W1shift}
 W_1(y;z;\psi;k) = \omega_1(y;\psi) + W_1^0(y;z;\psi;k),
\end{equation}
where the $z$-independent field $\omega_1(y)$ serves as the generating function for the HS gauge fields of the pure Chern-Simons HS theory \cite{Blencowe:1988gj}, while the conventional solution $W_1^0$ can be expressed as
\begin{equation}\label{W1^0shift}
W_1^0 = \frac{1}{4 i} \left[W_{0}*C*\Delta_{p+t}\Delta_{p} \gamma + C * W_{0}*\Delta_{p+2t}\Delta_{p+t} \gamma \right]\, .
\end{equation}

The decomposition into topological and dynamical fields for the one-forms is opposite to that for the zero-forms \eqref{C^dt}: $\omega\left(y ; \psi ; k\right) = \omega^{\text{dyn}}\left(y ; \psi_{1} ; k\right) + \omega^{\text{top}}\left(y ; \psi_{1} ; k\right) \psi_2$.
An analogous complementary picture for one- and zero-forms occurs in the $4d$ HS theory \cite{Vasiliev:1988sa}.

Substituting the expression \eqref{W1shift} into the HS Eqs. {\eqref{PV1}} yields the linearized equations of the form
\begin{equation}
 D_0 \omega_1 = - D_0 W_1^0,
\label{Xeqn37-37}
\end{equation}
where $D_0$ denotes the full covariant derivative \eqref{D0}. In the dynamical sector, the \rhs of this equation is trivial,
\begin{equation}\label{D0w1dyn}
 D_0 \omega_1^{\text{dyn}} (y;\psi_1) = 0 \,
\end{equation}
in agreement with the fact that $3d$ massless HS gauge fields do not propagate, being of Chern-Simons type (unlike the $AdS_4$ theory, where this step leads to non-zero $C$-dependent terms).

However, the equation for $\omega_1^\text{top}$ acquires a
$C^{\text{dyn}}$-dependent contribution,
\begin{equation}\label{D0w1top}
 D_0 \omega_1^\text{top} (y;\psi_1) =-\left.\frac{\lambda^2}{16i} h_{0}{ }^{\alpha}{}_{\gamma} h_{0}{}^{\gamma \beta}(y_\alpha-p_\alpha)(y_\beta-p_\beta) C^{\text{dyn}}\left(\xi ; \psi_{1}\right)\right|_{\xi=0}\neq 0.
\end{equation}
This equation establishes the dependence of the topological field $\omega_1^\text{top}$ on the dynamic field $C^{\text{dyn}}$. Therefore $\omega_1^\text{top}$ cannot be chosen freely. The related contribution can in principle induce nonlocal corrections. In HS theory, a natural setup is of spinor spin-locality \cite{Gelfond:2018vmi} requiring a finite number of spinor derivatives ${\partial}/{\partial y^\alpha}$, which can be related to space-time spin-locality \cite{Vasiliev:2022med}. However, the entanglement of the form \eqref{D0w1top} may require an explicit solution of the differential equation leading to a genuine space-time non-locality. The latter can manifest itself at higher-orders upon $\omega_1^\text{top}$ is substituted.

In \cite{Vasiliev:1992ix}, an appropriate field redefinition $\omega_1^\text{top} \rightarrow \omega_1^\text{top} + \delta \omega_1^\text{top}$, which compensates the \rhs\,\,of (\ref{D0w1top}), was found by hand,
\begin{equation}\label{dwold}
 \delta \omega_1^\text{top} (y) = -\frac{\lambda}{8i} \int_{0}^{1} d s\left(1-s^{2}\right) h_{0}^{\alpha \beta}(y_\alpha-p_\alpha)(y_\beta-p_\beta) C^{\text{dyn}}(s y) \psi_1 .
\end{equation}
Clearly, this is equivalent to adding a $\dr_z$-closed term to the solution of \eqref{dZW1}
$$W_1 = \omega_1 + W_1^0 + \delta \omega_1^\text{top}\psi_2.$$

The goal of \cite{Korybut:2022kdx} was to find disentangling solutions by the shifted homotopy approach.

\subsection{Advancements of the shifted homotopy approach}\label{Xsec5-1.4}

The shifted homotopy approach developed in \cite{Didenko:2018fgx} provides more freedom in the description of possible solutions to the entanglement problem. In \cite{Korybut:2022kdx}, a solution to \eqref{dZW1} was sought in the form of a sum of the conventional one and a correction derived through the shifted homotopy procedure:
\begin{equation}\label{W1shift}
W_1(y;z;\psi;k) = \omega_1(y;\psi) + W_1^0(y;z;\psi;k) + \delta W_1^\text{shift} (y;\psi;k).
\end{equation}

The correction $\delta W_1^\text{shift}$ was governed by six homotopy parameters: associated with the shift in the variable $y_\alpha$ and the derivatives $p_\alpha$ and $t_\alpha$ \eqref{dpt} of the fields $C$ and $W_0$, respectively. These were doubled since the \rhs of \eqref{dZW1} contains two $\dr_z$-closed terms associated with the different order of the one- and zero-forms. The requirement of triviality of the \rhs of the equation in the topological sector of $D_0 \omega_1 \equiv 0$ leaves only two free parameters associated with the $y_\alpha$-shifts. As shown in \cite{Korybut:2022kdx}, the resulting disentangling shifted homotopy correction is
\begin{equation}\label{dW1shift}
\begin{aligned}
\delta W_1^\text{shift} = -\frac{\lambda}{4 i} &\Big[h_{0}\psi_1 * C^\text{dyn}\psi_2 * h_{\alpha_+ y} \Delta_{p+t} \Delta_{p} \gamma + C^\text{dyn}\psi_2 * h_0\psi_1 * h_{2p+2t+\alpha_- y} \Delta_{p+2t} \Delta_{p+t} \gamma \Big]
=\\=
-\frac{\lambda}{8 i} \bigg[
\mathrlap{\int_0^1}  \quad ds &\cdot (1-s^2) h_0^{\alpha \beta} \big(
-y_\alpha p_\beta + p_\alpha p_\beta + \alpha_+ (y_\alpha y_\beta - p_\alpha p_\beta) \big) \psi_1 C^\text{dyn}(s y) \psi_2
+\\+
\mathrlap{\int_0^1}  \quad ds &\cdot (1-s^2) h_0^{\alpha \beta} \big(
+y_\alpha p_\beta - p_\alpha p_\beta + \alpha_- (y_\alpha y_\beta - p_\alpha p_\beta) \big) \psi_1 C^\text{dyn}(-s y) \psi_2
\bigg] k\,.
\end{aligned}
\end{equation}

Though describing a family of disentangling corrections with two free parameters $\alpha_\pm \in \mathds{C}$, this expression does not include the correction found in \cite{Vasiliev:1992ix}. To see this, one can calculate
\begin{equation}\label{dW1-dw1shift}
    \begin{gathered}
        \delta W_1^\text{shift} - \delta \omega_1^\text{top} \psi_2 = - \frac{\lambda}{8i} \int_0^1 ds \, (1-s^2)  h_0^{\alpha \beta}
        \times
        \\
        \begin{aligned}
            \times
            \Big[
            \big( \alpha_+ - \frac{1}{2} \big) (y_{\alpha}y_{\beta} - p_{\alpha} p_{\beta}) C^\text{dyn}(s y) &+
            \big( \alpha_- + \frac{1}{2} \big) (y_{\alpha}y_{\beta} - p_{\alpha} p_{\beta}) C^\text{dyn}(-s y)
            - \\ -
            (y_{\alpha}y_{\beta} + p_{\alpha} p_{\beta}) C_{+}^\text{dyn}(s y) &+
            2 y_{\alpha} p_{\beta} C_{-}^\text{dyn}(s y)
            \Big] \psi_1 \psi_2 k\,,
        \end{aligned}
    \end{gathered}
\end{equation}
where $C_{+}^{\text{dyn}}(s y)$ and $C_{-}^{\text{dyn}}(s y)$ denote the even and odd parts of $C^{\text{dyn}}(s y)$, respectively.
In \cite{Korybut:2022kdx}, it was shown that terms of the form $h_0^{\alpha \beta}(y_{\alpha} y_{\beta} - p_{\alpha} p_{\beta})$ in \eqref{dW1-dw1shift} are $D_0$-exact, while the other two are in $D_0$-cohomology\footnote{$D_0$-cohomology is understood as cohomology in the space of formal series in spinor variables, while from the $X$-space perspective we are interested in the local cohomology of $\dr_x$.}. Denoting them as ${A}_\pm$ and ${G}_\pm$, respectively, one can write
\begin{equation}\label{dWshiftstruct}
 \delta W_1^\text{shift} = \delta \omega_1^\text{top} \psi_2 + (\alpha_+ - \frac{1}{2}) {A}_+ + (\alpha_- + \frac{1}{2}) {A}_- + {G}_+ + {G}_-.
\end{equation}
From \eqref{dWshiftstruct}, it is clear that the shifted correction cannot reproduce the correction of \cite{Vasiliev:1992ix} by adjusting the free parameters.

The $D_0$-cohomological terms were shown in \cite{Korybut:2022kdx} to correspond to the mass deformation in $3d$ HS theory \cite{Korybut:2022kdx,Barabanshchikov:1996mc}. However, it was unclear within the shifted homotopy approach how to reproduce the full variety of the cohomological terms, which lead to disentangled equations.

The aim of this paper is to apply the recently developed differential homotopy procedure \cite{Vasiliev:2023yzx} to the description of disentangling solutions at the linear level of 3d HS theory. This approach allows one to make homotopy parameter-dependent shifts, postponing the integration to the very last step of the evaluation of corrections. The differential homotopy approach of \cite{Vasiliev:2023yzx} leads to a new representation for the homotopy solution of nonlinear HS equations, which reduces its analysis to a certain polyhedra cohomology problem with respect to the homotopy parameters. This technique allows us to represent the \rhs of \eqref{dzf} in the manifestly exact form, for which it was previously necessary to account for integration by parts and the Schouten identity.

In \cite{Vasiliev:2023yzx} (see also \cite{Kirakosiants:2025gpd}), the efficiency of the differential homotopy method has been demonstrated by constructing projectively-compact spin-local solutions with the minimal number of derivatives (for details, see \cite{Vasiliev:2022med}) in the second order of the $4d$ HS-theory. In this paper, it will be shown how it allows one to describe a set of disentangling solutions of the linear $3d$ HS theory, which contains all known solutions as well as some new.

\section{Differential homotopy}\label{Xsec6-2} \label{Diffhom}

\subsection{Total differential and integration}\label{Xsec7-2.1}

\setcounter{equation}{0}

In \cite{Vasiliev:2023yzx}, homotopy parameters such as $\tau$ or others that may enter through a homotopy shift $q^\alpha$ in \eqref{Dq} are denoted as a single set of variables $t^a$.
Along with the spinor variables $z^\alpha$, $t^a$ are interpreted as coordinates of the extended manifold $ M = \mathcal{M}_z \times \mathcal{M} $ with the total differential
\begin{equation}\label{drtot}
\dr := \dr_{z} + \dr_{t}, \quad \dr_{z} := \theta^\alpha \frac{\partial}{\partial z^\alpha}, \quad \dr_{t} := d t^{a} \frac{\partial}{\partial t^{a}},
\end{equation}
where $ \theta^\alpha $ and $ d t^{a} $ are anticommuting differentials, $ \mathcal{M}_z = \mathbb{R}^2 $ and $ \mathcal{M} $ is compact .

At each perturbation order, the equations to be solved take the form
\begin{equation}\label{drf}
\dr f(z, t, \theta, d t) = g(z, t, \theta, d t),
\quad \mbox{with} \quad
\dr g(z, t, \theta, d t) = 0.
\end{equation}
The homotopy coordinates $t^a(a=1, \ldots, n)$ vary in a bounded domain $\mathcal{M}$ of $\mathbb{R}^n$. Functions like $f, g$ must be integrable on $ \mathcal{M} $ and involve $\theta$- and $\delta$-functions such as $\theta(t) \theta(1-t)$ which restrict the integration to some compact domain.

Physical fields and equations are governed by the $\dr$-cohomology represented by $z^\alpha$-, $\theta^\alpha$-independent integrals over $ \mathcal{M} $.
The differential homotopy approach eliminates the integrals from the equations
\begin{equation}\label{fgint}
\dr_{z} f_{int} = g_{int}, \quad g_{int} = \int_{\mathcal{M}} g(z, \theta, t, d t), \quad f_{int} = \int_{\mathcal{M}} f(z, \theta, t, d t)
\end{equation}
leading to
\begin{equation}
\dr_{z} f = g + \dr_{t} h + g^{weak},
\label{Xeqn48-48}
\end{equation}
where the degree of $g^{weak}$ in $\mathcal{M}$ differs from $\dim \mathcal{M}$, so it does not contribute to the integral (for details, see \cite{Vasiliev:2023yzx}).
In this framework, Eqs. {\eqref{drf}} are invariant under diffeomorphisms of the total space $M$. Sometimes, it is also useful to extend this setup by introducing additional coordinates, such as $u^\alpha$ and $v^\alpha$, associated with the non-compact non-commutative space \eqref{*-prod}:
\begin{equation}\label{druv}
\dr:=\dr_{z}+\dr_{t}+\dr_{u}+\dr_{v}+\ldots, \quad \dr_{u}:=d u^{\alpha} \frac{\partial}{\partial u^{\alpha}}, \quad \dr_{v}:=d v^{\alpha} \frac{\partial}{\partial v^{\alpha}}.
\end{equation}

To avoid a sign ambiguity due to the (anti)commutativity of differential forms, we follow the conventions of \cite{Vasiliev:2023yzx},
\begin{equation}
\int_{t^{n}} \int_{t^{m}}=-\int_{t^{m}} \int_{t^{n}}
\quad , \quad
\int_{t^{1}} \cdots \int_{t^{k}} \rightarrow \int_{t^{1} \ldots t^{k}},
\label{Xeqn50-50}
\end{equation}
where $\int_{t^1 \ldots t^k}$ is totally antisymmetric in $t^a$. Also, we use the convention that $\int_{t^{1} \ldots t^{k}}$ can be written anywhere in the expression for the differential form to be integrated, with the convention
\begin{equation}
\dr \int_{t^{1} \ldots t^{k}} = (-1){}^{k} \int_{t^{1} \ldots t^{k}} \dr, \quad \lambda \int_{t^{1} \ldots t^{k}} = (-1){}^{k p} \int_{t^{1} \ldots t^{k}} \lambda,\qquad \forall \quad\mbox{p-form}\quad \lambda.
\label{Xeqn51-51}
\end{equation}

\subsection{Fundamental Ansatz}\label{Xsec8-2.2}

As demonstrated in \cite{Vasiliev:2023yzx}, direct perturbative analysis of nonlinear HS equations at lower orders yields expressions for fields of the form
\begin{equation}\label{ansatz}
f_{\mu}=\mathrlap{\int_{p_{i}^{2} r_{i}^{2} u^{2} v^{2} \tau \sigma \beta \rho}} \quad \mu(\tau, \sigma, \beta, \rho, u, v, p, r) \dr \Omega^{2} \mathcal{E}(\Omega) G_l(g(r)) \,,
\end{equation}
where $
G_{l}(g):=g_{1}\left(r_{1}\right) \ldots g_{l}\left(r_{l}\right) k\,
$
with the Klein operator $k$ \eqref{comm}, $g_{i}(y)$ are some functions of $y_{\alpha}$ (e.g., $C(y)$ or $\omega(y)$),
\begin{equation}\label{dOmega}
\dr \Omega^{2}:=\dr \Omega^{\alpha} \dr \Omega_{\alpha},
\end{equation}
\begin{equation}\label{Omega}
\Omega_{\alpha}(\tau, \sigma, \beta, \rho):=\tau z_{\alpha}-(1-\tau)\left(p_{\alpha}(\sigma)-\beta v_{\alpha}+\rho\left(y_{\alpha}+p_{+\alpha}+u_{\alpha}\right)\right),
\end{equation}
\begin{equation}
p_{+\alpha}:=\sum_{i=1}^{l} p_{i \alpha} ,\quad
p_{\alpha}(\sigma):=\sum_{i=1}^{l} p_{i \alpha} \sigma_{i} ,
\label{Xeqn55-55}
\end{equation}
with some integration parameters $\sigma_{i}$,
\begin{equation}\label{EOmega}
\mathcal{E}(\Omega):=\exp i\Big(\Omega_{\beta}\big(y^{\beta}+p_{+}^{\beta}+u^{\beta}\big)+u_{\alpha} v^{\alpha}-\sum_{i=1}^{l} p_{i \alpha} r_{i}^{\alpha}-\sum_{l \geq j>i \geq 1} p_{i \beta} p_{j}^{\beta}\Big).
\end{equation}

We will assume in the sequel that
\begin{equation}
d^{2} u := d u^{\alpha} d u_{\alpha}, \quad d^{2} v := d v^{\alpha} d v_{\alpha}, \quad d^{2} p_{i} := d p_{i}^{\alpha} d p_{i \alpha}, \quad d^{2} r_{i} := d r_{i}^{\alpha} d r_{i \alpha} ,
\label{Xeqn57-57}
\end{equation}
where the anticommuting differentials $d u^{\alpha}, d v^{\alpha}, d p_{i}^{\alpha}$, and $d r_{i}^{\alpha}$ are treated on the same footing as $\theta^{\alpha}, d \tau, d \sigma_{i}, d \rho$, and $d \beta$. In analogy with \eqref{druv}, we take
\begin{equation}\begin{split}
\dr = \dr_{z} + d \tau \frac{\partial}{\partial \tau} + d \rho \frac{\partial}{\partial \rho} + d \sigma_{i} \frac{\partial}{\partial \sigma_{i}} + d \beta \frac{\partial}{\partial \beta} + d u^{\alpha} \frac{\partial}{\partial u^{\alpha}} + d v^{\alpha} \frac{\partial}{\partial v^{\alpha}} + d p_{i}^{\alpha} \frac{\partial}{\partial p_{i}^{\alpha}} + d r_{i}^{\alpha} \frac{\partial}{\partial r_{i}^{\alpha}} ,
\label{Xeqn58-58}
\end{split}
\end{equation}
\begin{equation}\label{mumes}
\mu(\tau, \sigma, \beta, \rho, u, v, p, r) = \mu(\tau, \sigma, \beta, \rho) \, d^{2} u \, d^{2} v \prod_{i=1}^{l} d^{2} p_{i} \, d^{2} r_{i} ,
\end{equation}
with the normalization convention
$
\int d^{2} u \, d^{2} v \, \exp \left( i u_{\beta} v^{\beta} \right) = 1 .
$
Note that the terms involving $d p_{i}^{\alpha}, d r_{i}^{\alpha}, d u_{\alpha}$, and $d v_{\alpha}$ do not contribute to $\dr \Omega^\alpha$ and are not even treated as weakly zero, since the measure factor $d^{2} u \, d^{2} v \prod_{i=1}^{l} d^{2} p_{i} \, d^{2} r_{i}$ is present at all stages.

Using the Taylor expansion
$g_{i}(v_{i}) = \left. \exp \left( v_{i}^{\alpha} \frac{\partial}{\partial y^{\alpha}} \right) g_{i}(y) \right|_{y=0}$,
expressions \eqref{ansatz}, \eqref{EOmega} can be rewritten in differential form by integrating out $r_{i}^{\alpha}$, leading to
\begin{equation}
p_{j \alpha} = -i \frac{\partial}{\partial r_{j}^{\alpha}}
\label{Xeqn60-60}
\end{equation}
evaluated at $r_{j}^{\alpha} = 0$ upon differentiation in agreement with \eqref{dpt}.

The parameters $\sigma_{i}, \beta$, and $\rho$, now treated as integration variables are related to the shift parameters of \cite{Didenko:2018fgx,Didenko:2019xzz}.
The relation can be made explicit \cite{Kirakosiants:2025gpd} by defining $\mu(\tau, \sigma, \beta, \rho)$
to contain factors like $\DD(\beta-\beta_0)$, where
\begin{equation}
\DD(a):=\dr a \delta(a) .
\label{Xeqn61-61}
\end{equation}

The most important feature of the Ansatz \eqref{ansatz} is that its $\Omega$-dependent part is d-closed due to the Schouten identity
$\dr (\dr \Omega^2 \mathcal{E}(\Omega)) = (\dr \Omega){}^3 \mathcal{E}(\Omega) \equiv 0.$
As a result, the differential $\dr$ acting on the integral function $f_{\mu}$ \eqref{ansatz} actually acts only on the measure $\mu(\tau, \sigma, \beta, \rho)$ in \eqref{mumes},
\begin{equation}\label{Schoutendiff}
\dr f_{\mu}=(-1){}^{N} f_{\dr \mu},
\end{equation}
where $N$ is the number of integration parameters $\tau, \sigma_{i}, \beta, \rho$. This reformulates the original homological problem, expressed in terms of $z_\alpha$ variables, into one defined on the space of the measure factors $\mu(\tau, \sigma, \beta, \rho)$.
By \eqref{Schoutendiff}, one obtains an equation of the type \eqref{drf},
\begin{equation}
f_{\dr \mu_{f}} = g_{\mu_{g}}
\, \Rightarrow \,
\dr \mu_{f} \cong \mu_{g},
\label{Xeqn63-63}
\end{equation}
where $\mu_f$ and $\mu_g$ are measures depending on the homotopy parameters $\tau, \sigma_{i}, \beta, \rho$, and $\cong$ indicates equality up to weak terms that vanish upon integration (for details, see \cite{Vasiliev:2023yzx,Kirakosiants:2025gpd}).
The integration domains in $f_{\dr \mu_{f}}$ and $g_{\mu_{g}}$ are the polyhedra in $\mathbb{R}^n$ that are defined by the $\delta$- and $\theta$-functions included in the measures $\dr \mu_{f}$ and $\mu_{g}$.

The requirement that the function $g$ in \eqref{drf} is $\dr$-closed implies that $\mu_{g}$ must also be weakly $\dr$-closed,
$
\dr \mu_{g} \cong 0 .
$
Usually, this leads to the relation
\begin{equation}\label{mesdiffeq}
\mu_{g} \cong \dr h_{g}
\end{equation}
allowing one to set
$
\mu_{f} = h_{g} .
$
It should be noted that the measures must be defined globally in $\mathbb{R}^n$. A good example of a problem of the type \eqref{mesdiffeq} is the Newton-Leibniz formula, where only the difference of $\delta$-functions can be integrated: $\DD(a)-\DD(a-1)=\dr \big( \theta(a)\theta(1-a) \big)$.

This way, the differential homotopy reduces the solution of Eq. {\eqref{drf}} to solving differential equations of the form \eqref{mesdiffeq} for some compact measure dependent on homotopy parameters.

\subsection{Star multiplication formulas}\label{Xsec9-2.3}

The star-exchange relations of \cite{Didenko:2018fgx}, which describe the effect of star multiplication by a $z$-independent function on the outcome of the application of the contracting homotopy operator, play a key role in the shifted homotopy analysis of HS theory. The differential homotopy approach \cite{Vasiliev:2023yzx} provides a useful analogue of these formulas.

The star products of the function $f_{\mu}$ \eqref{ansatz} with a $z$-independent function $\varphi(y)$ yield
\begin{equation}\label{phi*F}
\varphi(y) * f_{\mu}=\mathrlap{\int_{p_{i}^{2} r_{i}^{2} u^{2} v^{2} \tau \sigma \beta \rho \rho \sigma_{\varphi}}} \quad \mathbb{D}\left(\sigma_{\varphi}+(1-\beta)\right) \mu(\tau, \sigma, \beta, p, r, u, v) \dr \Omega^{2} \mathcal{E}(\Omega) G_{k+1}(\varphi, g) \,
\end{equation}
with $\Omega_{\alpha}$ of the form \eqref{Omega} where $p_{+\alpha}$ now includes $p_{0 \alpha} \equiv p_{\varphi\alpha}$ and
\begin{equation}\label{F*phi}
f_{\mu} * \varphi(y)=\mathrlap{\int_{p_{i}^{2} r_{i}^{2} u^{2} v^{2} \tau \sigma \beta \rho \rho \sigma_{\varphi}}} \quad \mathbb{D}\left(\sigma_{\varphi}-(1-\beta)\right) \mu(\tau, \sigma, \beta, \rho, p, r, u, v) \dr \Omega^{2} \mathcal{E}(\Omega) G_{k+1}(g, \varphi)
\end{equation}
with $\Omega_{\alpha}$ \eqref{Omega} where $p_{k+1, \alpha} \equiv p_{\varphi \alpha}$.

When the star product with a $z$-independent $\varphi(y)$ acts on one of the inner factors of $g_{l}(y)$,
\begin{equation}
g_{l}(y) \rightarrow g_{l}(y) * \varphi(y) \quad \mbox{or}\quad
g_{l+1}(y) \rightarrow \varphi(y) * g_{l+1}(y) \,,
\end{equation}
one finds \cite{Vasiliev:2023yzx}
\begin{equation}\label{g*phi}
f_{\mu}\left(g_{l} * \varphi\right)=\mathrlap{\int_{\tau \sigma \beta \rho p_{i}^{2} r_{i}^{2} u^{2} v^{2} \sigma_{\varphi}}} \quad \DD(\sigma_\varphi-\sigma_{l}) \mu(\tau, \sigma, \beta, \rho, p, r, u, v) \dr \Omega^{2} \mathcal{E}(\Omega) G\left(\ldots g_{l}\left(r_{l}\right), g_{\varphi}\left(r_{\varphi}\right), g_{l+1}\left(r_{l+1}\right) \ldots\right) \,,
\end{equation}
\begin{equation}
\label{phi*g}
f_{\mu}\left(\varphi * g_{l+1}\right)=\mathrlap{\int_{\tau \sigma \beta \rho p_{i}^{2} r_{i}^{2}, u^{2} v^{2} \sigma_{\varphi}}} \quad \DD(\sigma_\varphi-\sigma_{l+1}) \mu(\tau, \sigma, \beta, \rho, p, r, u, v) \dr \Omega^{2} \mathcal{E}(\Omega) G\left(\ldots g_{l}\left(r_{l}\right), g_{\varphi}\left(r_{\varphi}\right), g_{l+1}\left(r_{l+1}\right) \ldots\right) \,.
\end{equation}

In accordance with \cite{Didenko:2019xzz}, under the condition $-\infty<\beta<1$, the variables $\sigma_{i}$ vary from $\beta-1$ to $1-\beta$,
\begin{equation}
\sigma_{i} \in[\beta-1,1-\beta] .
\label{Xeqn70-70}
\end{equation}
The formulas \eqref{phi*F}, \eqref{F*phi}, \eqref{g*phi}, and \eqref{phi*g} play an instrumental role in the perturbative analysis via
\begin{equation}\label{starleft}
\varphi(y) * f_{\mu}(g_1 \dots)-f_{\mu}\left(\varphi * g_1 \dots\right )=\mathrlap{\int_{\tau \sigma \beta \rho \sigma_{\varphi}}} \quad \dr P(\beta-1, \sigma_{\varphi}, \sigma_1) \mu(\tau, \sigma, \beta, \rho) \dr \Omega^{2} \mathcal{E}(\Omega) G\left(\varphi, g_1 \dots\right),
\end{equation}
\begin{equation}\label{starright}
f_{\mu}\left(\dots g_{k} * \varphi\right)-f_{\mu}(\dots g_{k}) * \varphi(y)=\mathrlap{\int_{\tau \sigma \beta \rho \sigma_{\varphi}}} \quad  \dr P(\sigma_{k}, \sigma_{\varphi}, 1-\beta) \mu(\tau, \sigma, \beta, \rho) \dr \Omega^{2} \mathcal{E}(\Omega) G\left(\ldots g_{k}, \varphi\right),
\end{equation}
where the spinor integration variables $p_{i}^{\alpha}, r_{i}^{\alpha}, u^{\alpha}$, and $v^{\alpha}$ are implicit, and
\begin{equation}\label{P2}
P\left(\sigma_{l}, \sigma_{l+1}, \sigma_{l+2}\right):=\theta\left(\sigma_{l+2}-\sigma_{l+1}\right) \theta\left(\sigma_{l+1}-\sigma_{l}\right)\,.
\end{equation}

The extension to an arbitrary string of arguments also plays a role in the analysis,
\begin{equation}\label{Pk}
P_{k}\left(\sigma_{l}, \sigma_{l+1}, \ldots, \sigma_{l+k}\right):=\theta\left(\sigma_{l+k}-\sigma_{l+k-1}\right) \ldots \theta\left(\sigma_{l+1}-\sigma_{l}\right) .
\end{equation}
The characteristic function $P(0, \sigma,1)$ of the unit segment $\sigma \in[0,1]$ will be denoted as $l(\sigma):=\theta(\sigma) \theta(1-\sigma) .$

\subsection{The differential homotopy form of the higher-spin equations}\label{Xsec10-2.4}

A notable innovation of the differential homotopy framework of \cite{Vasiliev:2023yzx} is its democratic treatment of space-time coordinates $x^\nu$, auxiliary spinor variables $z^\alpha$, homotopy parameters $t^{a}$, and even non-commutative star-product coordinates $u^\alpha$ and $v^\alpha$.
Accordingly, alongside the space-time differential $\dr_{x} := d x^\nu \frac{\partial}{\partial x^\nu}$, one introduces the total exterior differential $\dr$ \eqref{drtot}.
This setup enables a reformulation of the field equations in which all integrations over the homotopy parameters $t^{a}$ and the coordinates $u^\alpha$ and $v^\alpha$ are postponed to the very last step. For brevity, integrations over $u^\alpha$ and $v^\alpha$ will be implicit in the sequel.

The reformulation within the differential homotopy framework requires a modification of the $3d$ HS equations along the lines of the $4d$ construction of
\cite{Vasiliev:2023yzx}. The key requirement is that integration over the homotopy parameters must result in the original system \eqref{PV1}-\eqref{PV5}. To achieve this, the field $B$ is extended to incorporate differentials over the homotopy parameters $d t^{\alpha}$, the $z^{\alpha}$-differentials $\theta^{\alpha}$, and the space-time differentials, which enter via the one-form $\omega$,
\begin{equation}\label{Bdiff}
\textstyle \mathcal{B} = \sum_{n \geq 0} B^{(n)} ,
\end{equation}
where $B^{(n)}$ contains $n$ powers of $\omega$:
\begin{equation}
B^{(n)} = \tilde{B}^{(n)} \, \omega^n .
\label{Xeqn76-76}
\end{equation}

Similarly, $W$ is extended by the components that are space-time differential forms of higher degrees. The field $\mathcal{W}$ is introduced in the form
\begin{equation}
\theta^\alpha z_\alpha+\mathcal{W}:=W+S .
\label{Xeqn77-77}
\end{equation}

Note that the newly introduced fields appear under the differential $\dr$ and hence do not contribute to the field equations at $z^{\alpha} = \theta^{\alpha} = 0$ after integration over the homotopy parameters. As a result, the following differential homotopy system is equivalent to the original one \eqref{PV1}--\eqref{PV5}:
\begin{equation} \label{PVdiffB}
\left( \dr_x - 2 i \, \dr \right) \mathcal{B} + [\mathcal{W}, \mathcal{B}]_* \cong 0 ,
\end{equation}
\begin{equation} \label{PVdiffW}
\left( \dr_x - 2 i \, \dr \right) \mathcal{W} + \mathcal{W} * \mathcal{W} \cong i \mathcal{B} * \gamma .
\end{equation}

\setcounter{equation}{0}

\section{Linear level computations}\label{Xsec11-3} \label{Linearlevel}

Apart from the explicit form of the $AdS_3$ connection $W_0$, the computation of vertices in $AdS_3$ HS theory within the differential homotopy approach
is essentially the same as in the holomorphic sector of the $4d$ theory considered in \cite{Vasiliev:2023yzx}.
As a vacuum solution to Eqs. {\eqref{PVdiffB}}-\eqref{PVdiffW}, we choose
\begin{equation}
\mathcal{B}_{0}=0, \quad \mathcal{W}_{0}=W_0(y, \psi | x),
\label{Xeqn80-80}
\end{equation}
where $W_0(y, \psi | x)$ describes a flat HS connection. In $AdS_3$, it takes the bilinear form \eqref{AdSback}.

\subsection{$\mathcal{B}_1$ and $S_1$}\label{Xsec12-3.1}

In the lowest order, the Eq. {\eqref{PVdiffB}} on $\mathcal{B}_{1}$ yields
\begin{equation}\label{D0B1}
D_0 \mathcal{B}_{1}-2 i \dr \mathcal{B}_{1}=0,\qquad
D_{0} = \dr_x + [W_{0}, ]_* .
\end{equation}
This implies that the $ d x $-independent part of $ \mathcal{B}_{1} $ is $ \dr $-closed. The gauge transformations of the form $\delta \Phi = \dr \varepsilon + \dots$ allow one to gauge away the $ \dr $-exact part (for details, see equations (5.13)-(5.15) in \cite{Vasiliev:2023yzx}). Choosing $ \mathcal{B}_{1} = C(y; \psi; k | x) $ in $ \dr $-cohomology $ H^{0}(\dr) $, the $ d x $-dependent part of Eq. {\eqref{D0B1}} yields the standard free HS Eq. {\eqref{D0C}}.

Eq. {\eqref{PVdiffW}} yields
\begin{equation}\label{dS1}
\begin{aligned}
-2 i \dr S_{1}
=
i&\mathrlap{\int_{\tau \rho \beta \sigma}} \quad \DD(1-\tau) \mu(\rho, \beta, \sigma) \dr \Omega^2 \mathcal{E}(\Omega) C(r_c) k
\cong \\ \cong
i\dr &\mathrlap{\int_{\tau \rho \beta \sigma}} \quad \theta(\tau) \theta(1-\tau) \mu(\rho, \beta, \sigma) \dr \Omega^2 \mathcal{E}(\Omega) C(r_c) k \,,
\end{aligned}
\end{equation}
with any $\dr$-closed measure $\mu(\rho, \beta, \sigma)$ normalized
to obey
\begin{equation} \label{munorm}
{\int_{\rho \beta \sigma}} \mu(\rho, \beta, \sigma)= 1 .
\end{equation}
The equivalence in \eqref{dS1} is up to the term with $\DD(\tau)$, which is weak.

In the $AdS_4$ case of \cite{Vasiliev:2023yzx}, this measure is fixed as $\mu(\rho, \beta, \sigma) = \DD(\sigma) \tilde{\mu}(\rho, \beta)$, since the $4d$ on-mass-shell theorem involves $ C(0) $. In the 3d case under consideration, the on-mass-shell theorem is trivial in the dynamical sector \eqref{D0w1dyn}, while the goal is to trivialize it in the topological one. Hence, at this stage, the measure $\mu(\rho, \beta, \sigma)$ may be chosen in its most general form, constrained only by \eqref{munorm}, which leads to
\begin{equation}\label{S1diff}
S_1 = - \frac{1}{2} \mathrlap{\int_{\tau \rho \beta \sigma}} \quad l(\tau) \mu(\rho, \beta, \sigma) \dr \Omega^2 \mathcal{E}(\Omega) C(r_c) k \,.
\end{equation}

\subsection{$W_1$ solution}\label{Xsec13-3.2}

The equation for $W_1$ resulting from \eqref{PVdiffW} has the form
\begin{equation}\label{drW1}
-2 i \dr W_1+D_0 S_1 \cong 0 .
\end{equation}
Using \eqref{starleft}-\eqref{starright}, one obtains
\begin{equation} \label{D0S1diff}
\begin{aligned}
D_0 S_1= - \frac{1}{2} \mathrlap{\int_{\tau \rho \beta \sigma \sigma_\omega}} \quad   l(\tau) \mu(\rho, \beta, \sigma) \dr \Omega^2 \mathcal{E}(\Omega) \dr \big[  &P(\beta-1, \sigma_\omega, \sigma) W_0(r_\omega) C(r_C)
+\\+
&P(\sigma, \sigma_\omega, 1-\beta)  C(r_C) W_0(r_\omega) \big] k\,.
\end{aligned}
\end{equation}
The general solution to \eqref{drW1} is
\begin{equation}\label{W1struct}
 W_1 = \omega_1 + W_1^\mu + \delta W_1 .
\end{equation}
As in \eqref{W1shift}, $\omega_1$ is the generating function for the HS gauge fields.
A particular solution $W_1^\mu$ of \eqref{drW1} with the measure $\mu(\rho, \beta, \sigma)$ from \eqref{S1diff} is
\begin{equation}\label{W1mu}
\begin{aligned}
W_1^\mu =
\frac{i}{4 } \mathrlap{\int_{\tau \rho \beta \sigma \sigma_\omega}} \quad   l(\tau) \mu(\rho, \beta, \sigma) \dr \Omega^2 \mathcal{E}(\Omega) \big[  &P(\beta-1, \sigma_\omega, \sigma) W_0(r_\omega) C(r_C)
+\\+
& P(\sigma, \sigma_\omega, 1-\beta)  C(r_C) W_0(r_\omega) \big] k\,.
\end{aligned}
\end{equation}

A correction $\delta W_1$ must be $\dr$-closed as a solution to the homogeneous equation $-2i \dr W_1 \cong 0$. Discarding $\dr$-exact terms that do not contribute upon integration, we demand $\delta W_1$ be in the $\dr$-cohomology, hence containing the measure factor of $\DD(\tau)$, which trivializes the $z$-dependence. As a result, due to the absence of $\theta_\alpha $ in $\dr \Omega_\alpha$ at $\tau=0$, $\dr (\delta W_1)$ is weakly zero. We consider $\delta W_1$ of the form close to $W_1^\mu$ with some two-form measure $\nu(\rho, \beta, \sigma)$,
\begin{equation}\label{dW1nu}
\begin{aligned}
\delta W_1^\nu =
\frac{1}{4 i}
\mathrlap{\int_{\tau \rho \beta \sigma \sigma_\omega}} \quad \DD(\tau) \nu(\rho, \beta, \sigma) \dr \Omega^2 \mathcal{E}(\Omega) \big[ &P(\beta-1,\sigma_\omega,\sigma) W_0(r_\omega) C(r_c) 
+\\+ 
&P(\sigma,\sigma_\omega,1-\beta) C(r_c) W_0(r_\omega) \big]k \,.
\end{aligned}
\end{equation}

\subsection{Equations of motion for $\omega_1$}\label{Xsec14-3.3} \label{HSVertex}

The linearized equation for the HS gauge fields generating function $\omega_1$ is derived from \eqref{PVdiffW}
\begin{equation} \label{D0W1}
D_0 \omega_1 + D_0 (W_1^\mu) + D_0 (\delta W_1^\nu) + \dr W_{1}^{(2)} \cong 0 ,
\end{equation}
where $W_{1}^{(2)}$ is a space-time two-form.
Using \eqref{starleft}-\eqref{starright}, it is not difficult to obtain
\begin{equation}
\begin{aligned}
&D_0 (W_1^\mu) =
- \frac{1}{4 i}
\mathrlap{\int_{\tau \rho \beta \sigma \sigma_{\omega_1} \sigma_{\omega_2}}} \quad
l(\tau) \mu(\rho, \beta, \sigma) \dr \Omega^2 \mathcal{E}(\Omega)
\times
\dr \big[  P_3(\beta-1, \sigma_{\omega_1}, \sigma_{\omega_2}, \sigma) W_0(r_{\omega_1}) W_0(r_{\omega_2}) C(r_c)
+\\+
&P_4(\beta-1, \sigma_{\omega_1}, \sigma, \sigma_{\omega_2}, 1-\beta) W_0(r_{\omega_1}) C(r_c) W_0(r_{\omega_2})
+
P_3(\sigma, \sigma_{\omega_1}, \sigma_{\omega_2}, 1-\beta) C(r_c) W_0(r_{\omega_1}) W_0(r_{\omega_2}) \big] k ,
\end{aligned}
\end{equation}

\begin{equation}
\begin{aligned}
&D_0 (\delta W_1^\nu) =
\frac{1}{4 i}
\mathrlap{\int_{\tau \rho \beta \sigma \sigma_{\omega_1} \sigma_{\omega_2}}} \quad
\DD(\tau) \nu(\rho, \beta, \sigma) \dr \Omega^2 \mathcal{E}(\Omega)
\times
\dr \big[ P_3(\beta-1, \sigma_{\omega_1}, \sigma_{\omega_2}, \sigma) W_0(r_{\omega_1}) W_0(r_{\omega_2}) C(r_c)
+\\+
&P_4(\beta-1, \sigma_{\omega_1}, \sigma, \sigma_{\omega_2}, 1-\beta) W_0(r_{\omega_1}) C(r_c) W_0(r_{\omega_2})
+
P_3(\sigma, \sigma_{\omega_1}, \sigma_{\omega_2}, 1-\beta) C(r_c) W_0(r_{\omega_1}) W_0(r_{\omega_2}) \big] k .
\end{aligned}
\end{equation}

From here, one can find the space-time two-form $W_{1}^{(2)}$ as
\begin{equation}
\begin{aligned}
&W_{1}^{(2)} =
\frac{1}{4 i}
\mathrlap{\int_{\tau \rho \beta \sigma \sigma_{\omega_1} \sigma_{\omega_2}}} \quad
\big( \DD(\tau) \nu(\rho, \beta, \sigma) - l(\tau) \mu(\rho, \beta, \sigma) \big)
\dr \Omega^2 \mathcal{E}(\Omega)
\big[ P_3(\beta-1, \sigma_{\omega_1}, \sigma_{\omega_2}, \sigma) W_0(r_{\omega_1}) W_0(r_{\omega_2}) C(r_c)
+\\&+
P_4(\beta-1, \sigma_{\omega_1}, \sigma, \sigma_{\omega_2}, 1-\beta) W_0(r_{\omega_1}) C(r_c) W_0(r_{\omega_2})
+
P_3(\sigma, \sigma_{\omega_1}, \sigma_{\omega_2}, 1-\beta) C(r_c) W_0(r_{\omega_1}) W_0(r_{\omega_2}) \big] k .
\end{aligned}
\end{equation}

By omitting the weak term with $\DD(1-\tau)$, we obtain the equation for $\omega_1$ in the form
\begin{equation}\label{Dw1diff}
\begin{aligned}
&D_0 \omega_1 =
\frac{1}{4 i}
\mathrlap{\int_{\tau \rho \beta \sigma \sigma_{\omega_1} \sigma_{\omega_2}}} \quad
\DD(\tau) \big( \mu(\rho, \beta, \sigma) + \dr \nu(\rho, \beta, \sigma) \big) \dr \Omega^2 \mathcal{E}(\Omega)
\big[ P_3(\beta-1, \sigma_{\omega_1}, \sigma_{\omega_2}, \sigma) W_0(r_{\omega_1}) W_0(r_{\omega_2}) C(r_c)
+\\+
&P_4(\beta-1, \sigma_{\omega_1}, \sigma, \sigma_{\omega_2}, 1-\beta) W_0(r_{\omega_1}) C(r_c) W_0(r_{\omega_2})
+
P_3(\sigma, \sigma_{\omega_1}, \sigma_{\omega_2}, 1-\beta) C(r_c) W_0(r_{\omega_1}) W_0(r_{\omega_2}) \big] k \,.
\end{aligned}
\end{equation}

By substituting the $AdS_3$ connection \eqref{AdSback} and expanding in powers of $\psi_2$ \eqref{C^dt}, one can see that, after integration over $\tau$, $\sigma_{\omega_1}$ and $ \sigma_{\omega_2}$, all terms containing $\omega_0$ or $C^\text{top}$ cancel. The \rhs of \eqref{Dw1diff} takes the form
\begin{equation}\label{D0w1exact}
\begin{aligned}
D_0 \omega_1 = - \frac{\lambda^2}{16i} \mathrlap{\int_{ \rho \beta \sigma}} \quad
\big( \mu(\rho, \beta, \sigma) + \dr \nu(\rho, &\beta, \sigma) \big) (1-\beta)^2 \left(1-\frac{\sigma^2}{(1-\beta)^2}\right)^2
\times \\ \times
&h_{0}{ }^{\alpha}{}_{\gamma}  h_{0}{}^{\gamma \beta}(y_\alpha-p_\alpha)(y_\beta-p_\beta) C^\text{dyn} \left(-\frac{\sigma y}{1-\beta}\right) \psi_2 k \,.
\end{aligned}
\end{equation}

The conventional solution leading to entangling of topological and dynamical fields \eqref{D0w1top} corresponds to the measures of the forms
\begin{equation}\label{mu0}
\mu_0(\rho, \beta, \sigma) = - \DD(\rho) \DD(\beta) \DD(\sigma), \qquad
\nu_0(\rho, \beta, \sigma) =0 .
\end{equation}

Thus, the problem of finding solutions that do not entangle topological and dynamical fields is reduced to selecting such measures $\mu(\rho, \beta, \sigma)$ and $\nu(\rho, \beta, \sigma)$ that the \rhs of \eqref{D0w1exact} is zero,
\begin{equation}\label{D0w1=0}
 D_0 \omega_1 \equiv 0 .
\end{equation}

Because the arguments of $h_0$ and $C^{\text{dyn}}$ on the \rhs of \eqref{D0w1exact} do not involve $\rho$, this variable can be integrated out. Since the field $C^{\text{dyn}}(y)$ is a formal power series \eqref{Cgen} in $y_\alpha$, the integral must be zero at each degree of $\left(-\frac{\sigma}{1-\beta}\right)$, that is, the integration measure must be such that
\begin{equation}\label{meas->0}
\mathrlap{\int_{ \rho \beta \sigma}} \quad
\big( \mu(\rho, \beta, \sigma) + \dr \nu(\rho, \beta, \sigma) \big) (1-\beta){}^2 \left(1-\frac{\sigma^2}{(1-\beta){}^2}\right){}^2 \left(-\frac{\sigma}{1-\beta}\right){}^n \equiv 0
\end{equation}
at any $n \in \mathds{N} \cup \{0\}$. In agreement with the results of \cite{Didenko:2019xzz}, $1-\beta>0$, so we can fix $\beta=0$ using the measure $\DD(\beta)$ and adjust only the $\sigma$ variable. Thus, the measure $\big( \mu(\rho, \beta, \sigma) + \dr \nu(\rho, \beta, \sigma) \big)$ in \eqref{D0w1exact} is required to contain $\DD(\beta) \DD(\sigma \pm 1)$.

\setcounter{equation}{0}

\section{Disentangling via $\delta W_1$}\label{Xsec15-4} \label{DisentanglingdW1}

In \cite{Korybut:2022kdx}, disentangling solutions were found by selection of a suitable correction $\delta W_1^\text{shift}$ \eqref{W1shift}. To reproduce the results of the shifted-homotopy approach using differential-homotopy, we fix the measure in \eqref{S1diff} in the conventional way as $\mu_0(\rho, \beta, \sigma) = - \DD(\rho) \DD(\beta) \DD(\sigma)$ and we will select a suitable measure $\nu(\rho, \beta, \sigma)$ in $\delta W_1^\nu$ that leads to the disentangled Eq. {\eqref{D0w1=0}}.

One possible form of the disentangling correction is
\begin{equation}\label{dW1diff}
\begin{aligned}
\delta W_1^{\nu_1} =
\frac{1}{4 i}
\mathrlap{\int_{\tau \rho \beta \sigma \sigma_\omega}} \quad \DD(\tau) \nu_1(\rho, \beta, \sigma) \dr \Omega^2 \mathcal{E}(\Omega) \big[ &P(\beta-1,\sigma_\omega,\sigma) W_0(r_\omega) C(r_c) 
+\\+ 
&P(\sigma,\sigma_\omega,1-\beta) C(r_c) W_0(r_\omega) \big]k \,,
\end{aligned}
\end{equation}
where the measure is
\begin{equation}
\nu_1(\rho, \beta, \sigma)= \dfrac{1}{2} \DD(\rho) \DD(\beta) (l(\sigma) - l(-\sigma)) .
\label{Xeqn100-100}
\end{equation}

The measure in \eqref{D0w1exact} corresponding to this correction has the form
\begin{equation}
\begin{gathered}
\mu_0(\rho, \beta, \sigma) + \dr \nu_1(\rho, \beta, \sigma) =
- \frac{1}{2} \DD(\rho) \DD(\beta) \left( \DD(\sigma+1) + \DD(\sigma-1) \right) ,
\end{gathered}
\label{Xeqn101-101}
\end{equation}
which satisfies the condition \eqref{meas->0}.

One can easily verify that the expression \eqref{dW1diff} corresponds to the shifted homotopy correction \eqref{dW1shift} with $\alpha_\pm = 0$,
\begin{equation}\label{dW1a1a20}
\begin{aligned}
\delta W_1^{\nu_1} = \left. \delta W_1^\text{shift} \right|_{\alpha_+=\alpha_-=0} =
-\frac{\lambda}{8 i} \bigg[
\mathrlap{\int_0^1}  \quad ds &\cdot (1-s^2) h_0^{\alpha \beta} (
-y_\alpha p_\beta + p_\alpha p_\beta ) \psi_1 C^\text{dyn}(s y) \psi_2
+\\+
\mathrlap{\int_0^1}  \quad ds &\cdot (1-s^2) h_0^{\alpha \beta} (
y_\alpha p_\beta - p_\alpha p_\beta ) \psi_1 C^\text{dyn}(-s y) \psi_2
\bigg] k \,.
\end{aligned}
\end{equation}

\subsection{$D_0$-exact corrections}\label{Xsec16-4.1} \label{D0-exactcorr}

The $D_0$-exact terms ${A}_\pm$ in \eqref{dWshiftstruct} and corresponding zero-forms $\epsilon_\pm$ such that $D_0 \epsilon_\pm = {A}_\pm$ are
\begin{equation}\label{Apm}
{A}_\pm =
-\frac{\lambda}{8i} \mathrlap{\int_0^1} \quad ds (1-s^2) h_0^{\alpha \beta} (y_\alpha y_\beta - p_\alpha p_\beta) C^\text{dyn}(\pm s y) \psi_1 \psi_2 k \,,
\qquad
\epsilon_\pm = -\frac{1}{4} \mathrlap{\int_0^1} \quad ds C(\pm s y)k \,.
\end{equation}
As shown in \cite{Korybut:2022kdx}, the expression $h_0^{\alpha \beta} (y_\alpha y_\beta - p_\alpha p_\beta) C^{\text{dyn}}(\pm s y) \psi_1 \psi_2 k$ is $D_0$-closed independently of the integration measure, as long as the integral converges. Thus, one can write a more general $D_0$-closed one-form ${A}^f_\pm$ with the corresponding zero-form $\epsilon^f_\pm$
\begin{equation}\label{Afexact}
{A}^f_\pm = -\frac{\lambda}{8i} \mathrlap{\int_0^1} \quad ds f(s) h_0^{\alpha \beta} (y_\alpha y_\beta - p_\alpha p_\beta) C^\text{dyn}(\pm s y) \psi_1 \psi_2 k \,,
\quad
\epsilon^f_\pm = -\frac{1}{4} \mathrlap{\int_0^1} \quad ds \frac{f(s)}{1-s^2} C(\pm s y)k \,.
\end{equation}
From the form of ${A}^f_\pm$, it follows that the only restriction on $f(s)$ is integrability on $[0,1]$. However, if $f(1) \neq 0$, then the zero form $\epsilon^f_\pm$, naively constructed by analogy with \eqref{Apm}, can be divergent. This raises the question of how to construct a convergent zero-form $\epsilon^f_\pm$ corresponding to the $D_0$-closed one-form ${A}^f_\pm$.

This problem can be solved by representing the $D_0$-exact forms in the differential homotopy approach. Initially, one must obtain the structure $ h_0^{\alpha \beta}(y_\alpha y_\beta - p_\alpha p_\beta) $ within the fundamental Ansatz. To this end, it is necessary to adjust the integration measure over the parameter $\rho$ associated with the variables $ y_\alpha$. Indeed, due to \eqref{dOmega}-\eqref{Omega}, the measure $\DD(\tau) \DD(\rho) \DD(\beta)$ in \eqref{dW1nu} leads only to the pre-exponential derivatives of the form $ t^\gamma p_\gamma$.
With the bilinear connection $ W_0(y) $ this can only induce linear combinations of the structures $ h_0^{\alpha \beta} y_\alpha p_\beta $ and $ h_0^{\alpha \beta} p_\alpha p_\beta $.

To obtain the structure $ h_0^{\alpha \beta} y_\alpha y_\beta $, one has to let $ \rho $ be nonzero, for instance, by using the measure $ \DD(\rho + \sigma) $, which introduces derivatives $ t^\gamma y_\gamma $ into the pre-exponential. This way, the $D_0 $-exact one-form $A^f_\pm$ from \eqref{Afexact} can be reproduced in the differential homotopy language as
\begin{equation} \label{Afpm}
\begin{aligned}
\mathcal{A}_\pm^f = \frac{1}{8i}
\mathrlap{\int_{\tau \rho \beta \sigma \sigma_\omega}} \quad \DD(\tau) \big( \DD(\rho) &- \DD(\rho+\sigma) \big) \DD(\beta) l(\mp\sigma) \frac{ f(\mp\sigma)}{1-\sigma^2} \dr \Omega^2 \mathcal{E}(\Omega) 
\times\\&\times
\big[ P(-1,\sigma_\omega,\sigma) W_0(r_\omega) C(r_c) + P(\sigma,\sigma_\omega,1) C(r_c) W_0(r_\omega) \big]k \,.
\end{aligned}
\end{equation}

The corresponding zero-form $\varepsilon_\pm^f$ such as $D_0 (\varepsilon_\pm^f) = \mathcal{A}_\pm^f$ has the following structure
\begin{equation} \label{epsfpm}
\varepsilon_\pm^f = \pm \frac{1}{8i} \mathrlap{\int_{\tau \rho \beta \sigma}} \quad \DD(\tau) \theta(\pm \rho) \theta(\mp(\sigma+\rho)) \DD(\beta) \theta(1\pm\sigma) F(\mp\sigma) \dr \Omega^2 \mathcal{E}(\Omega) C(r_c)k \,,
\end{equation}
where the function $F(-\sigma)$ is related to $f(s)$ in \eqref{Afexact} via the differential equation
\begin{equation}\label{dsF(s)}
(1-s^2) \frac{d}{ds}\big(s F(s)\big) = f(s) .
\end{equation}
This equation can be solved in the form
\begin{equation}\label{F(s)}
F(s)=\frac{\text{const}}{s}+\frac{1}{s} \int_0^s \frac{f(\xi)}{1-\xi^2} d\xi .
\end{equation}

Note that, due to the measure $\theta(\pm \rho) \theta(\mp(\sigma+\rho))$ in \eqref{epsfpm}, the integration over $\rho$ gives $\mp \sigma$, which eliminates the pole at $\sigma=0$. This measure, originally introduced to ensure the structure $h_0^{\alpha \beta}(y_\alpha y_\beta - p_\alpha p_\beta)$ in \eqref{Afpm}, automatically provides integrability in \eqref{epsfpm}.

It is easy to check that $\mathcal{A}_\pm^f$ \eqref{Afpm} reproduces the one-form $A^f_\pm$ \eqref{Afexact} for $AdS_3$ connection $W_0(y)$ \eqref{AdSback}.
However, the convergence requirement affects the zero-form.
Substituting \eqref{F(s)} into \eqref{epsfpm} and integrating over $\tau$ and $\beta$ in both terms, and also integrating over $\rho$ in the term with $f(\xi)$, one obtains two converging integrals
\begin{equation}\label{epstwoterms}
\begin{aligned}
\varepsilon_\pm^f = \pm \frac{1}{4i} \bigg[
&\mathrlap{\int_{\sigma \xi}} \quad d\xi d\sigma \theta(1 \pm \sigma) \theta(\mp \sigma-\xi) \theta(\xi) \frac{f(\xi)}{1-\xi^2}  y^\gamma p_\gamma C(-\sigma y)k
+\\+
&\mathrlap{\int_{\sigma \rho}} \quad d\rho d\sigma \theta(1 \pm \sigma) \theta(\mp (\sigma+\rho)) \theta(\pm \rho) \frac{\text{const}}{\mp \sigma}  y^\gamma p_\gamma C(-\sigma y)k \bigg] \,.
\end{aligned}
\end{equation}

Note that here $y^\gamma p_\gamma=i\frac{\partial}{\partial(-\sigma)}$, so one can integrate by parts
\begin{equation} \label{epsfpm'}
\varepsilon_\pm^f = -\frac{1}{4} {\int_0^1} ds \frac{f(s)}{1-s^2} C(\pm s y)k
+ \frac{1}{4} F(1) C(\pm y)k .
\end{equation}
Since $D_0C(\pm y)=0$, it differs from $\epsilon_\pm^f$ in \eqref{Afexact} by a $D_0$-closed term with $F(1)$. As is natural to expect, the integration constant of \eqref{F(s)} contributes only via $F(1)$ in the $D_0$-closed term.

Notably, if $\dfrac{f(s)}{1-s^2}$ has a pole at $s=1$, then $F(s)$ also has a pole, and the $D_0$-closed term $F(1) C(\pm y)$ effectively compensates for the divergence in the integral. This mechanism of the divergence compensation would be difficult to guess without using the differential homotopy approach.
Note that this phenomenon is somewhat analogous to the problem of reproducing the $\delta B_2$ shift leading to a vertex with the minimal number of derivatives in the $4d$ HS theory \cite{Vasiliev:2016xui}, which was reproduced within the differential homotopy approach in \cite{Vasiliev:2023yzx}.

Thus, the differential homotopy approach makes it possible to reproduce the general form of $D_0$-exact expressions \eqref{Afpm}.
In particular, the differential homotopy approach allows one to describe
the family of solutions obtained earlier using shifted homotopy \eqref{dW1shift} as
\begin{equation}\label{dW1_shift_diff}
\begin{gathered}
\delta W_1^\text{shift} = \delta W_1^{\nu_1} + \mathcal{A}_+^{\alpha_+ (1-s^2)} + \mathcal{A}_-^{{\alpha_- (1-s^2)}}
=\\
\begin{aligned}
=-\frac{1}{8 i}
\mathrlap{\int_{\tau \rho \beta \sigma \sigma_\omega}} \quad \DD(\tau) \Big((1-\alpha_+)\DD(\rho) + \alpha_+ \DD(\rho+\sigma) \Big) \DD(\beta) l(-\sigma) \dr \Omega^2 \mathcal{E}(\Omega) &\times \\ \times \big[ P(\beta-1,\sigma_\omega,\sigma) W_0(r_\omega) C(r_c) + P(\sigma,\sigma_\omega,1-\beta) &C(r_c) W_0(r_\omega) \big]k
+\\+
\frac{1}{8 i}
\mathrlap{\int_{\tau \rho \beta \sigma \sigma_\omega}} \quad \DD(\tau) \Big((1+\alpha_-)\DD(\rho) - \alpha_- \DD(\rho+\sigma) \Big) \DD(\beta) l(\sigma) \dr \Omega^2 \mathcal{E}(\Omega) &\times \\ \times \big[ P(\beta-1,\sigma_\omega,\sigma) W_0(r_\omega) C(r_c) + P(\sigma,\sigma_\omega,1-\beta) &C(r_c) W_0(r_\omega) \big]k \,.
\end{aligned}
\end{gathered}
\end{equation}

\subsection{Old solution}\label{Xsec17-4.2}

Once the role of the parameter $\rho$ as generating the $y_\alpha$-dependent pre-exponential terms has been clarified, the old correction \eqref{dwold} can be reproduced within the differential homotopy approach. The difference between the expressions \eqref{dwold} and \eqref{dW1a1a20} is that it contains the structure $h_0^{\alpha \beta} y_\alpha y_\beta$ and is free of the terms with $C(-s y)$. As a result, $\delta \omega_1^\text{top}$ can be written in the form
\begin{equation}\label{dwolddiff}
\begin{aligned}
\delta \omega_1^\text{top} \psi_2 = -\frac{1}{8i} \mathrlap{\int_{\tau \rho \beta \sigma \sigma_\omega}} \quad \DD(\tau) \left( \DD(\rho) + \DD(\rho+\sigma) \right) \DD(\beta) l(-\sigma) \dr \Omega^2 \mathcal{E}(\Omega)\big[
&P(\beta-1,\sigma_\omega,\sigma) W_0(r_\omega) C(r_c)
+\\+
&P(\sigma,\sigma_\omega,1-\beta) C(r_c) W_0(r_\omega) \big]k \,.
\end{aligned}
\end{equation}
It is easy to check that the substitution of the $AdS_3$ connection \eqref{AdSback} into \eqref{dwolddiff} yields \eqref{dwold}.

Here, the advantages of the differential homotopy approach can be appreciated: to find a solution of the required form or compare two different solutions, it is enough to work with the integral measures. For instance, the difference \eqref{dW1-dw1shift} between $\delta W_1^\text{shift}$ and $\delta \omega_1^\text{top}$ reads as
\begin{equation}\label{dW1-dw1diff}
\begin{gathered}
\delta W_1^{shift} - \delta \omega_1^\text{top} \psi_2 = \\
\begin{aligned}
=\frac{1}{8 i}
\mathrlap{\int_{\tau \rho \beta \sigma \sigma_\omega}} \quad \DD(\tau) &\Big((\alpha_+\DD(\rho) + (1-\alpha_+) \DD(\rho+\sigma) \Big) \DD(\beta) l(-\sigma) \dr \Omega^2 \mathcal{E}(\Omega) \times \\ &\times \big[ P(\beta-1,\sigma_\omega,\sigma) W_0(r_\omega) C(r_c) + P(\sigma,\sigma_\omega,1-\beta) C(r_c) W_0(r_\omega) \big]k
+\\+
\frac{1}{8 i}
\mathrlap{\int_{\tau \rho \beta \sigma \sigma_\omega}} \quad \DD(\tau) &\Big((1+\alpha_-)\DD(\rho) - \alpha_- \DD(\rho+\sigma) \Big) \DD(\beta) l(\sigma) \dr \Omega^2 \mathcal{E}(\Omega) \times \\ &\times \big[ P(\beta-1,\sigma_\omega,\sigma) W_0(r_\omega) C(r_c) + P(\sigma,\sigma_\omega,1-\beta) C(r_c) W_0(r_\omega) \big]k \,.
\end{aligned}
\end{gathered}
\end{equation}
In this formalism, the free parameters $\alpha_\pm$ appear explicitly as factors, {\it i.e.}, not via homotopy operators $h_q$ and $\Delta_q$ as in the shifted homotopy formalism.

\subsection{$D_0$-cohomology}\label{Xsec18-4.3}

Let us recall that cohomological terms in \eqref{dW1-dw1shift}, \eqref{dWshiftstruct} are expressed via even $C_{+}^{\text{dyn}}(s y)$ and odd $C_{-}^{\text{dyn}}(s y)$ parts of the function $C^{\text{dyn}}(s y)$,
\begin{equation}\label{G+}
{G}_+ = \frac{\lambda}{8i} \int_0^1 ds \, (1-s^2) h_0^{\alpha \beta} (y_{\alpha}y_{\beta} + p_{\alpha} p_{\beta}) C_{+}^{\text{dyn}}(s y) \psi_1 \psi_2 k,
\end{equation}
\begin{equation}\label{G-}
{G}_- = - \frac{\lambda}{4i} \int_0^1 ds \, (1-s^2) h_0^{\alpha \beta} y_{\alpha} p_{\beta} C_{-}^{\text{dyn}}(s y) \psi_1 \psi_2 k,
\end{equation}

One can easily reproduce the concrete $D_0$-cohomology $\mathcal{G}_{coh} = {G}_+ + {G}_-$ in differential homotopy form. Based on the form of the expression \eqref{dWshiftstruct}, one just has to substitute $\alpha_+ = -\alpha_- = \frac{1}{2}$ into \eqref{dW1-dw1diff},
\begin{equation}\label{Gcoh}
\begin{gathered}
\mathcal{G}_{coh} = {G}_+ + {G}_- =
\big( \delta W_1^{shift} - \delta \omega_1^\text{top} \psi_2 \big) \Big|_{\alpha_+=-\alpha_-=\frac{1}{2}} = \\
\begin{aligned}
=\frac{1}{16 i}
\mathrlap{\int_{\tau \rho \beta \sigma \sigma_\omega}} \quad \DD(\tau) &\Big( \DD(\rho) + \DD(\rho+\sigma) \Big) \DD(\beta) P(-1,\sigma,1) \dr \Omega^2 \mathcal{E}(\Omega)
\times \\ &\times
\big[ P(-1,\sigma_\omega,\sigma) W_0(r_\omega) C(r_c) + P(\sigma,\sigma_\omega,1) C(r_c) W_0(r_\omega) \big] k \,.
\end{aligned}
\end{gathered}
\end{equation}
Note that the explicit expressions in \eqref{G+}, \eqref{G-} are $D_0$-closed due to the fact that the expressions under the integral are even with respect to $s$. In differential homotopy representation \eqref{Gcoh} the expression is also even in $\sigma$ since all odd parts are $\dr$-exact.

Since the cohomologies ${G}_+$ and ${G}_-$ are independent, it is interesting to understand how each of them can be reproduced separately.
This problem is more tricky since integration over $\sigma_\omega$ with the measures $P(-1,\sigma_\omega,\sigma)$ and $P(\sigma,\sigma_\omega,1)$ for bilinear $W_0(y)$ \eqref{AdSback} yields $h_0^{\alpha\beta}(y-p)_\alpha \, (\ldots)_\beta$. Neither \eqref{G+} nor \eqref{G-} can be represented in such a form. Thus, it is necessary to modify the integration measure over $\sigma_\omega$.

In terms of differential homotopy, the appropriate representation of the $D_0$-cohomologies $G_-$ and $G_+$ \eqref{G+}-\eqref{G-} extended to arbitrary flat connection $W_0(y)$ is
\begin{equation}\label{Gpmdiff}
\begin{aligned}
\mathcal{G}_\pm = \frac{1}{16i}
\mathrlap{\int_{\tau \rho \beta \sigma \sigma_\omega}} \quad \DD(\tau) \DD(\rho + \frac{1\pm1}{2} \sigma) \DD(\beta) \dr \Omega^2 \mathcal{E}(\Omega) \Big[ l(-\sigma) \frac{1}{2m_y} P(a_y,\sigma_\omega,b_y) W_0(r_\omega) C(r_c&)
+\\+
l(\sigma) \frac{1}{2n_y} P(c_y,\sigma_\omega,d_y) C(r_c) W_0(r_\omega&) \Big] k
+\\+\frac{1}{16i}
\mathrlap{\int_{\tau \rho \beta \sigma \sigma_\omega}} \quad \DD(\tau) \DD(\rho + \frac{1\mp1}{2} \sigma) \DD(\beta) \dr \Omega^2 \mathcal{E}(\Omega) \Big[ l(-\sigma) \frac{1}{2m_p} P(a_p,\sigma_\omega,b_p) W_0(r_\omega) C(r_c&)
+\\+
l(\sigma) \frac{1}{2n_p} P(c_p,\sigma_\omega,d_p) C(r_c) W_0(r_\omega&) \Big] k \,,
\end{aligned}
\end{equation}
where $m_y, n_y, m_p, n_p$ are some real positive numbers, and integration limits are
\begin{equation}
\left\{
\begin{aligned}
a_y &= -1-m_y+(1-m_y)\sigma\\
b_y &= -1+m_y+(1+m_y)\sigma \\
c_y &= 1-n_y+(1+n_y) \sigma \\
d_y &= 1+n_y+(1-n_y)\sigma \\
\end{aligned}
\right.
\qquad \quad ; \quad \qquad
\left\{
\begin{aligned}
a_p &= -m_p(1+\sigma) \\
b_p &= m_p(1+\sigma) \\
c_p &= -n_p(1-\sigma) \\
d_p &= n_p(1-\sigma) \\
\end{aligned}
\right. \, \quad .
\end{equation}

For the $AdS_3$ connection \eqref{AdSback} the expressions for $\mathcal{G}_\pm$ \eqref{Gpmdiff} are reduced to $G_\pm$ \eqref{G+}-\eqref{G-}.
Note that $\mathcal{G}_+ + \mathcal{G}_-$ is equal to $\mathcal{G}_{coh}$ \eqref{Gcoh} also only for the bilinear connection $W_0(y)$, since the integration measures in \eqref{Gcoh} and \eqref{Gpmdiff} have different structures.

Therefore, the differential homotopy approach now allows one to describe all solutions that disentangle dynamical and topological fields in terms of various solutions to Eq. {\eqref{drW1}},
\begin{equation}\label{W1full}
 W_1 = \omega_1+W_1^{\mu_0} + \delta W_1^{\nu_1} + \mathcal{A}_+^{f_+} + \mathcal{A}_-^{f_-} + \gamma_+ \mathcal{G}_+ + \gamma_- \mathcal{G}_- ,
\end{equation}
where $W_1^{\mu_0}$ is the conventional solution \eqref{W1mu} with the measure $\mu_0$ \eqref{mu0}, $\delta W_1^{\nu_1}$ is the disentangling correction \eqref{dW1diff}, $\mathcal{A}_\pm^{f_\pm}$ are $D_0$-exact terms \eqref{Afpm} with some measure functions $f_\pm (s)$, and $\mathcal{G}_\pm$ are $D_0$-cohomological terms \eqref{Gpmdiff} with two free parameters $\gamma_\pm \in \mathds{R}$.

\setcounter{equation}{0}

\section{Disentangling via $S_1$}\label{Xsec19-5} \label{DisentanglingS1}

Apart from the method of deriving disentangled equations via corrections to
the field $W_1$ \eqref{drW1}, as in \cite{Korybut:2022kdx}, relation \eqref{D0w1exact} suggests an alternative method via unconventional solution for $S_1$ in \eqref{dS1}. Namely, let the correction $\delta W_1$ in \eqref{W1struct} be $D_0$-closed being a combination of the $D_0$-exact \eqref{Afpm} and $D_0$-cohomological \eqref{Gpmdiff} terms. Then, for the condition \eqref{meas->0}, the measure in the solution for $S_1$ \eqref{S1diff} can be chosen in the form
\begin{equation} \label{munonconv}
\mu_1(\rho, \beta, \sigma) = \tilde{\mu}(\rho) \DD(\beta) \DD(\sigma \pm 1) .
\end{equation}
Note that this choice of the measure implies that $S_1$ is a function of $(y^\alpha \pm z^\alpha)$.

To establish the relation between the disentangling solutions obtained via the choice of $\delta W_1$ and that resulting from the choice of $S_1$ with the measure \eqref{munonconv} consider the difference between $S_1$ with the measure $\mu_1$ and the conventional one with $\mu_0$ \eqref{mu0},
\begin{equation}
\delta S_1 = S_1^{\mu_1}-S_1^{\mu_0} = - \frac{1}{2} \mathrlap{\int_{\tau \rho \beta \sigma}} \quad l(\tau) (\mu_1-\mu_0)(\rho, \beta, \sigma) \dr \Omega^2 \mathcal{E}(\Omega) C(r_c) k \,.
\end{equation}

That $S_1^{\mu_1}$ and $S_1^{\mu_0}$ both solve \eqref{dS1} implies $\dr (\delta S_1) = 0$.
It is not difficult to see that $\delta S_1$ is also $\dr$-exact by solving the equation
\begin{equation} \label{mu1-mu0}
\mu_1(\rho, \beta, \sigma)-\mu_0(\rho, \beta, \sigma) = \dr \nu (\rho, \beta, \sigma) .
\end{equation}
Indeed, normalization \eqref{munorm} for \eqref{munonconv} leads to $\int_\rho \tilde{\mu}(\rho)=-1$, so $\tilde{\mu}(\rho)$ can be chosen as $\tilde{\mu}(\rho)=-\DD(\rho)+\dr \varepsilon(\rho)$, where $\varepsilon(\rho)$ is some zero-form with compact support. Then the solution to \eqref{mu1-mu0} has the form
\begin{equation}
\nu (\rho, \beta, \sigma) = - \DD(\rho) \DD(\beta) \theta(\mp \sigma) \theta(1 \pm \sigma) + \varepsilon(\rho) \DD(\beta) \DD(\sigma \pm1) ,
\label{Xeqn123-123}
\end{equation}
and, up to weak terms, $\delta S_1=\dr \epsilon_\nu$, where
\begin{equation}
\epsilon_\nu = \frac{1}{2} \mathrlap{\int_{\tau \rho \beta \sigma}} \quad l(\tau) \nu (\rho, \beta, \sigma) \dr \Omega^2 \mathcal{E}(\Omega) C(r_c) k \,.
\end{equation}

For conventional $S_1^{\mu_0}$, the disentangling solution $W_1$ has the structure
\begin{equation} \label{W1convS1}
W_1 = \omega_1 + W_1^{\mu_0} + \delta W_1^\nu + \mathcal{A}_\pm^{f_\pm} + \gamma_\pm \mathcal{G}_\pm ,
\end{equation}
where $W_1^{\mu_0}$ has the form \eqref{W1mu}, $\delta W_1^\nu$ is some disentangling correction of the form \eqref{dW1nu}, and the remaining terms are $D_0$-closed.

For nonconventional $S_1^{\mu_1}$ with the measure \eqref{munonconv}, there is no need for a disentangling correction, so the structure of disentangling solution $W_1'$ is
\begin{equation}
W_1' = \omega_1 + W_1^{\mu_1} + \mathcal{A}_\pm^{f'_\pm} + \gamma'_\pm \mathcal{G}_\pm .
\label{Xeqn126-126}
\end{equation}

Note that the Eq. {\eqref{drW1}} with nonconventional $S_1^{\mu_1}$ can be written as
\begin{equation}
\dr W_1' = \frac{1}{2i} D_0 S_1^{\mu_1} = \frac{1}{2i} D_0 S_1^{\mu_0} + \frac{1}{2i} D_0 (\delta S_1) = \frac{1}{2i} D_0 S_1^{\mu_0} - \frac{1}{2i} \dr \left( D_0 \epsilon_\nu\right) ,
\label{Xeqn127-127}
\end{equation}
where the exact term can be moved to the left, leading to the equation
\begin{equation}
\dr \left(W_1' + \frac{1}{2i} D_0 \epsilon_\nu \right) = \frac{1}{2i} D_0 S_1^{\mu_0} ,
\label{Xeqn128-128}
\end{equation}
whose disentangling solution is \eqref{W1convS1}. Therefore, apart from the terms $\mathcal{A}_\pm^{f_\pm}$ and $\mathcal{G}_\pm$, solutions with different choices of the measure in $S_1$ can be related as
\begin{equation}
W_1^{\mu_0} + \delta W_1^\nu = W_1^{\mu_1} + \frac{1}{2i} D_0 \epsilon_\nu,
\label{Xeqn129-129}
\end{equation}
where the measures $\mu_1$, $\mu_0$, and $\nu$ are connected via \eqref{mu1-mu0}.

\section{Conclusion}\label{Xsec20-6} \label{Conclusion}

In this paper, the recently developed differential homotopy approach
\cite{Vasiliev:2023yzx} is applied to the derivation of the form of linearized equations of the HS theory in $AdS_3$. The advantage of this approach in which all expressions are represented as integrals over some polyhedra in a multidimensional space is that it allows one to avoid accounting for the Schouten identity.
The differential homotopy approach turns out to be both more general and simpler than the shifted homotopy approach of \cite{Didenko:2018fgx}.

The new method allowed us to obtain a general form of field equations with
disentangled sectors of dynamical and topological fields. This general solution includes both those obtained within the framework of shifted homotopy in \cite{Korybut:2022kdx} and those derived by hand in \cite{Vasiliev:1992ix}, which was unreachable within the shifted homotopy technique. Also, within the differential homotopy approach, a more general form of $D_0$-exact corrections was found, as well as $D_0$-cohomological terms identified in \cite{Korybut:2022kdx}. We anticipate that the described family contains all possible $D_0$-exact and $D_0$-cohomological corrections to the solution of $3D$ linear equations. In particular, this is supported by the fact that the found cohomologies were related to mass corrections of the $3d$ matter fields induced by the deformed oscillator algebra \cite{Barabanshchikov:1996mc}.

Also, within the differential homotopy framework, an alternative way to derive the disentangled equations through non-conventional solutions for the field $S_1$ is suggested.

The obtained results are important for further analysis of nonlinear
corrections to $3d$ HS field equations and, in particular, of their locality
properties, which may depend on the choice of a particular disentangling
solution. It is anticipated that the broad choice of disentangling solutions of the linearized equations will allow us to choose the most appropriate one for the
next order analysis.
As a first step, it is planned to evaluate local current interactions in the $3d$ HS theory.

\section*{Acknowledgments}
The authors thank Anatoly Korybut for careful reading the text and valuable comments and Philipp Kirakosiants for fruitful discussion of the results of the work.
MV is grateful for hospitality to Ofer Aharony,
Theoretical High Energy Physics Group of Weizmann Institute of Science where a
part of this work has been done.
This work was supported by Theoretical Physics and Mathematics Advancement Foundation “BASIS” Grant No 24-1-1-9-5.

\end{document}